\documentclass[twocolumn,10pt, twoside]{IEEEtran}
\usepackage{epsfig,amsfonts,subfigure,color}
\usepackage[nolist]{acronym}
\usepackage{graphicx,cite,amssymb,amsmath,mathrsfs}
\usepackage{verbatim}
\usepackage{hyperref}
\usepackage{tikz,amsmath, amssymb,bm,color}
\usepackage{verbatim}
\usepackage{mathrsfs}
\usepackage{amsthm}
\usepackage{balance}

\usepackage{pgfplots,relsize}

\newtheorem{theorem}{Remark}

\newcommand{\cer}[1]{$\mathrm{CER}\approx 10^{#1}$}

\hypersetup{
	colorlinks=true,       
	linkcolor=black,          
	citecolor=black,        
	filecolor=magenta,      
	urlcolor=blue           
}

\begin{document}
\begin{acronym}
\acro{AR3A}{accumulate-repeat-3-accumulate}
\acro{ARJA}{accumulate-repeat-jagged-accumulate}
\acro{AWGN}{additive white Gaussian noise}
\acro{BCH}{Bose-Chaudhuri-Hocquenghem}
\acro{bi-AWGN}{binary-input additive white Gaussian noise}
\acro{CCSDS}{Consultative Committee for Space Data Systems}
\acro{CER}{codeword error rate}
\acro{EXIT}{extrinsic information transfer}
\acro{FFT}{fast Fourier transform}
\acro{LDPC}{low-density parity-check}
\acro{ML}{maximum likelihood}
\acro{OSD}{ordered statistics decoding}
\acro{PEG}{progressive edge growth}
\acro{RA}{random access}
\acro{RCB}{random coding bound}
\acro{SC}{successive cancellation}
\acro{SNR}{signal-to-noise ratio}
\acro{SPB}{sphere packing bound}
\acro{WAVA}{wrap-around Viterbi algorithm}
\end{acronym}
\title{Code Design for Short Blocks: A Survey}
\author{Gianluigi Liva, Lorenzo Gaudio, Tudor Ninacs and Thomas Jerkovits
	
	\thanks{G. Liva, L. Gaudio, T. Ninacs and T. Jerkovits are with Institute of Communication and Navigation of the Deutsches Zentrum
		f\"{u}r Luft- und Raumfahrt (DLR), 82234 Wessling,
		Germany (e-mail: {\tt gianluigi.liva@dlr.de}).
	}
    \thanks{A preliminary version of this work was presented at the 25th Edition of the European Conference on Networks and Communications (EuCNC), June 2016.}
}

\date{\today}
\thispagestyle{empty} \setcounter{page}{1} \maketitle

\begin{abstract}
The design of block codes for short information blocks (e.g., a thousand or less information bits) is an open research problem which is gaining relevance thanks to emerging applications in wireless communication networks. In this work, we review some of the most recent code constructions targeting the short block regime, and we compare then with both finite-length performance bounds and classical error correction coding schemes. We will see how it is possible to effectively  approach the theoretical bounds, with different performance vs. decoding complexity trade-offs.
\end{abstract}

\section{Introduction}\label{sec:intro}

\IEEEPARstart{D}{uring} the past sixty years, a formidable effort  has been channeled in the research of capacity-approaching error correcting codes \cite{Shannon48}.
Initially the attention was directed to short and medium-length linear block codes \cite{berlekamp74:key} (with some notable exceptions, see e.g. \cite{Elias54:EFC,Gallager63:LDPC}), mainly for complexity reasons.
As the idea of code concatenation \cite{Forney66:Concatenated} got established in the coding theorists community \cite{Costello07:PROC}, the design of long channel codes became a viable solution to approach channel capacity. The effort resulted in a number of practical code constructions allowing reliable transmission at fractions of decibels from the Shannon limit \cite{Berrou93:TC,MacKay99:LDPC,Richardson01:Design,richardson01:capacity,Luby01:LDPC,Pfister05:capacity,Pfister07:ARA,Arikan09:Polar,Lentmaier10:CLDPC,Kudekar11:CLDPC} with low-complexity (sub-optimum) decoding.

The interest in short and medium-block length codes (i.e., codes with dimension $k$ in the range of $50$ to $1000$ bits) has been rising again recently, mainly due to emergent applications requiring the transmission of short data units. Examples of such applications are machine-type communications, smart metering networks, remote command links and messaging services (see e.g. \cite{Decola11:CCSDS,Boccardi14:MAG,Paolini15:MAG,Durisi16:Short}).

When the design of short iteratively-decodable codes is attempted, it  turns out that some classical code construction tools which have been developed for turbo-like codes tend to fail in providing codes with acceptable performance. This is the case, for instance, of density evolution \cite{Richardson08:BOOK} and \ac{EXIT} charts \cite{tenBrink01:EXIT}, which are well-established techniques to design powerful long \ac{LDPC} and turbo codes. The issue is due to the asymptotic (in the block length) nature of density evolution and \ac{EXIT} analysis which fail to properly model the iterative decoder in the short block length regime. However, competitive \ac{LDPC} and turbo code designs for moderate-length and small blocks have been proposed, mostly based on heuristic construction techniques \cite{Sadjadpour00:shortTC,Koetter03:Wheel,Ryan04:eIRA,Liva05:Tanner_Milcom,Divsalar07:ShortProto,Liva08:QC_GLDPC,Bocharova09:shortLDPC,Divsalar14:ProtoRaptor,Jerkovits16:Short,Davey98:NonBinary,Berkmann00:Diss,Berkmann02:dualtrellis,Poulliat08:BinImag,Venkiah08:RPEG,Chen09:Hamilton,Costantini10:NBLDPC,kasai11:multiplicatively,Liva12:IRAq,Divsalar12:NonBinaryShort,Liva13:ShortTC,Matuz13:LRNBTC,Dolecek14:NBLDPCTIT}.
While iterative codes retain a large appeal due their low decoding complexity, more sophisticated decoding algorithms \cite{Fossorier95:OSD,Fossorier04:BOX,Wu07:MRB,Liva14:OptProd,Tal15:ListPolar} are feasible for short blocks leading to solutions that are performance-wise competitive (if not superior) with respect to iterative decoding of short turbo and \ac{LDPC} codes.\footnote{Further approaches deserving a particular attention for short and moderate-length codes are, among others, those in \cite{Huber07:Multibase,Huber10:Multibase}.}

\section{A Case Study}\label{sec:case}

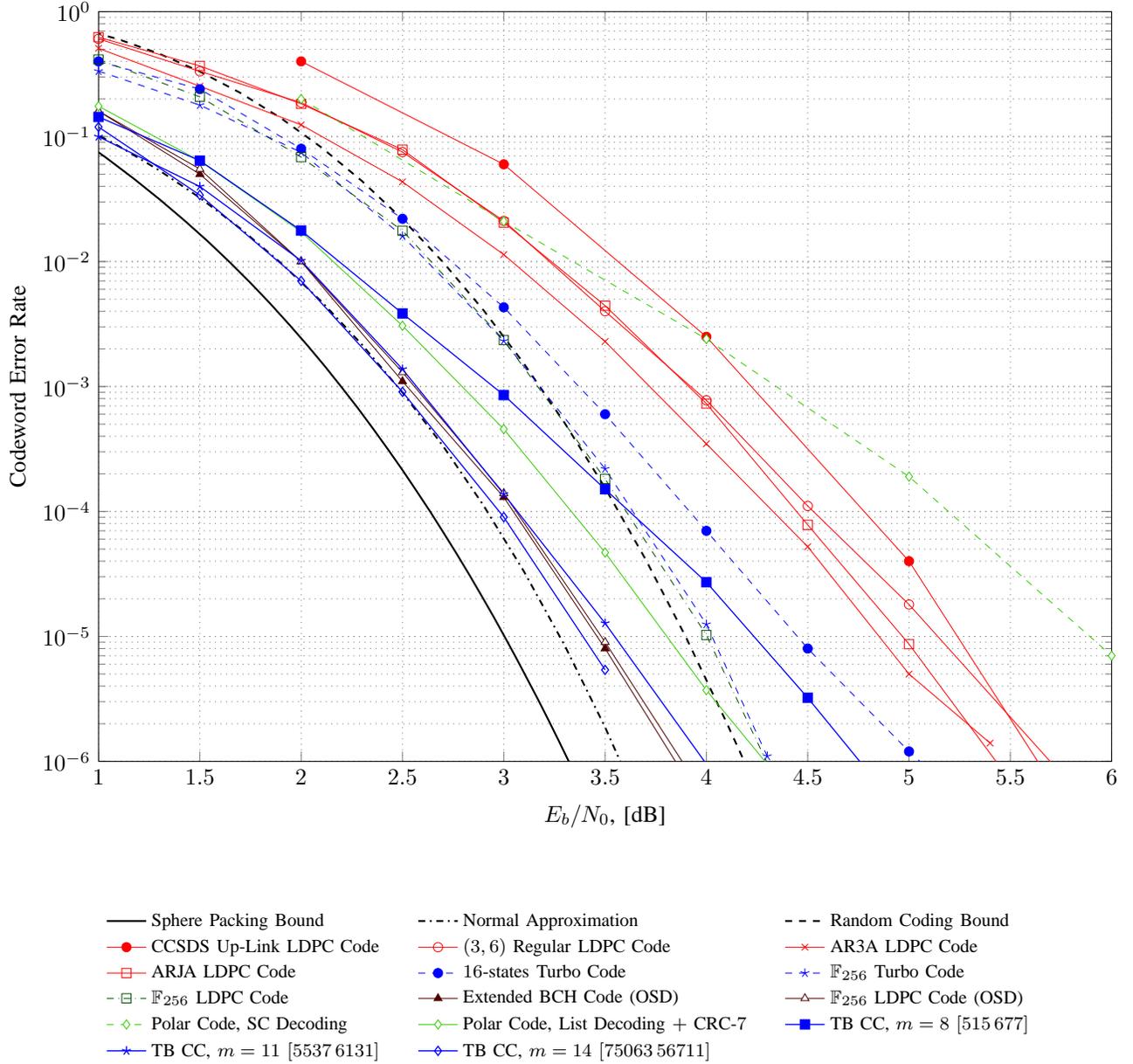
\begin{figure*}[t]
	\begin{center}
		\begin{tikzpicture}
\begin{semilogyaxis}[
     width=17cm,
     height=13cm,
     grid=both,
     grid style={dotted,gray},
       legend cell align=left,
       legend style={font=\footnotesize},
       legend columns=3,
       legend style={at={(0,-0.3)},anchor=west,draw=none,
       	/tikz/every even column/.append style={column sep=0.5cm}
       },
	xmin=1,
	xmax=6,
	ymin=1e-6,
	ymax=1,
	xlabel={$E_b/N_0$, [dB]},
	ylabel={Codeword Error Rate}
	]
\addplot[color=black,line width = 0.8pt,solid] table[x=snr,y=cer] {./spb_k64_n128.txt}; \label{leg:spb} \addlegendentry{Sphere Packing Bound};
\addplot[color=black,line width = 0.8pt,dashdotted] table[x=snr,y=cer] {./pvp_k64_n128.txt}; \label{leg:pvp} \addlegendentry{Normal Approximation};
\addplot[color=black,line width = 0.8pt,dashed] table[x=snr,y=cer] {./rcb_k64_n128.txt}; \label{leg:rcb} \addlegendentry{Random Coding Bound};


\addplot[color=red,line width = 0.3pt,solid,mark=*] table[x=snr,y=cer] {./protonasa.txt};\label{lab:protonasa} \addlegendentry{CCSDS Up-Link LDPC Code};
\addplot[color=red,line width = 0.3pt,solid,mark=o] table[x=snr,y=cer] {./gallager.txt}; \label{lab:36} \addlegendentry{$(3,6)$ Regular LDPC Code};
\addplot[color=red,line width = 0.3pt,solid,mark=x] table[x=snr,y=cer] {./ARA.txt}; \label{lab:ara}  \addlegendentry{AR3A LDPC Code};
\addplot[color=red,line width = 0.3pt,solid,mark=square] table[x=snr,y=cer] {./ARJA.txt}; \label{lab:arja} \addlegendentry{ARJA LDPC Code};


\addplot[color=blue,line width = 0.3pt,dashed,mark=*,mark color=blue,mark options={solid,fill=blue}] table[x=snr,y=cer] {./turbo16.txt}; \label{lab:turbo} \addlegendentry{$16$-states Turbo Code};


\addplot[color=blue,line width = 0.3pt,dashed,mark=star,,mark options={solid}] table[x=snr,y=cer] {./tc256.txt}; \label{lab:tc256} \addlegendentry{$\mathbb F_{256}$ Turbo Code};
\addplot[color=green!30!black,line width = 0.3pt,dashdotted,mark=square,,mark options={solid}] table[x=snr,y=cer] {./ldpc256.txt}; \label{lab:ldpc256} \addlegendentry{$\mathbb F_{256}$ LDPC Code};


\addplot[color=red!30!black,line width = 0.3pt,solid,mark=triangle*,mark options={solid,fill=red!30!black}] table[x=snr,y=cer] {./bch.txt}; \label{lab:bch} \addlegendentry{Extended BCH Code (OSD)};
\addplot[color=red!30!black,line width = 0.3pt,solid,mark=triangle*,mark options={solid,fill=white}] table[x=snr,y=cer] {./ldpcosd.txt}; \label{lab:ldpcosd} \addlegendentry{$\mathbb F_{256}$ LDPC Code (OSD)};


\addplot[color=green!80!red,line width = 0.3pt,dashed,mark=diamond*,mark options={solid,fill=white}] table[x=snr,y=cer] {./polarSC.txt}; \label{lab:polar} \addlegendentry{Polar Code, SC Decoding};
\addplot[color=green!80!red,line width = 0.3pt,solid,mark=diamond*,mark options={solid,fill=white}] table[x=snr,y=cer] {./polarList_CRC7.txt}; \label{lab:polar_list_crc} \addlegendentry{Polar Code, List Decoding $+$ CRC-$7$};


\addplot[color=blue,line width = 0.5pt,solid,mark=square*] table[x=snr,y=cer] {./k64_n128_TBm8.tik}; \label{lab:tb8} \addlegendentry{TB CC, $m=8$ $[515\, 677]$};
\addplot[color=blue,line width = 0.5pt,solid,mark=star] table[x=snr,y=cer] {./k64_n128_TBm11.tik}; \label{lab:tb11} \addlegendentry{TB CC, $m=11$ $[5537\, 6131]$};
\addplot[color=blue,line width = 0.5pt,solid,mark=diamond] table[x=snr,y=cer] {./tb14.txt}; \label{lab:tb14} \addlegendentry{TB CC, $m=14$ $[75063 \, 56711]$};

\end{semilogyaxis}
\end{tikzpicture}
	\end{center}
	\caption{Codeword Error Rate for several $(128,64)$ codes over the \ac{bi-AWGN} channel.}\label{fig:64128all}
\end{figure*}

In this section, we provide an exemplary comparison of short codes. We focus on the case study of codes with block length and code dimension $n=128$ and $k=64$ bits, respectively, which are the parameters of the shortest code recently standardized by \ac{CCSDS} \cite{CCSDS} for satellite telecommand links \cite{CCSDS14}. The performance of the schemes is measures in terms of \ac{CER} versus \ac{SNR} over the \ac{bi-AWGN} channel, with \ac{SNR} given by the $E_b/N_0$ ratio (here, $E_b$ is the energy per information bit and $N_0$ the single-sided noise power spectral density). Besides, we discuss other metrics such as the capability to detect errors and (although not exhaustively) the complexity of decoding.
For this block size, we defined a list of viable candidate solutions comprising
\begin{itemize}
\item[i.] Short binary \ac{LDPC} and turbo codes, and their non-binary counterparts;
\item[ii.] The $(128,64)$ extended \ac{BCH} code (with minimum distance $22$), under \ac{OSD};
\item[iii.] Two tail-biting convolutional codes with memory $m=8$ and $m=11$;
\item[iv.] A polar code under \ac{SC} decoding and under CRC-aided list decoding.
\end{itemize}
The performance of the codes is compared in Figure \ref{fig:64128all} with three finite-length performance benchmarks, i.e., the 1959 Shannon's \ac{SPB}\footnote{Additionally to Shannon's 1959 \ac{SPB}, one may consider the comparison with bounds relying on error exponents following the 1967 \ac{SPB} \cite{shannon1967lower,valembois2004sphere,wiechman2008improved}.} \cite{Shannon59:SPB} (\ref{leg:spb}), Gallager's \ac{RCB} \cite{Gallager68:BOOK} for the \ac{bi-AWGN} channel (\ref{leg:rcb}), and the normal approximation of \cite{Polyanskiy10:BOUNDS} (\ref{leg:pvp}).\footnote{Excellent surveys on performance bounds in the finite block length regime are given in \cite{Dolinar98:BOUNDS,Polyanskiy10:BOUNDS,SasonShamai06:BOUNDS}. A useful library of routines for the calculation of the benchmarks is available at \url{https://sites.google.com/site/durisi/software} \cite{SPECTRE}.} As reference, the performance of the $(128,64)$ binary protograph-based \cite{Thorpe03:PROTO,Divsalar07:ShortProto} \ac{LDPC} code from the \ac{CCSDS} telecommand standard \cite{CCSDS14} is provided too (\ref{lab:protonasa}). The \ac{CCSDS} \ac{LDPC} code performs somehow poorly in terms of coding gain. The code is outperformed at moderate error rates (\cer{-4}) even by a standard regular $(3,6)$ \ac{LDPC} code (\ref{lab:36}). The \ac{CCSDS} \ac{LDPC} is also outperformed by an \ac{AR3A} \ac{LDPC} code \cite{Divsalar04:ARA} (\ref{lab:ara}) an by an \ac{ARJA} \ac{LDPC} code \cite{Divsalar09:ProtoJSAC} (\ref{lab:arja}).\footnote{All \ac{LDPC} codes introduced in this comparison have been designed through a girth optimization technique based on the \ac{PEG} algorithm \cite{Hu05:PEG}. A maximum of $200$ belief propagation iterations have been used in the simulations (though, the average iteration count is much lower, especially at high \acp{SNR} thanks to early decoding stopping rules).} At low error rates (e.g. \cer{-6}) the \ac{CCSDS} \ac{LDPC} code is likely to attain lower error rates than the above-introduced  \ac{LDPC} code competitors thanks to its remarkable distance properties \cite{Divsalar07:ShortProto}. The four binary \ac{LDPC} codes introduced so far perform relatively poorly with respect to the benchmarks (roughly $1$ dB away from the \ac{RCB} at \cer{-4}). Despite its uninspiring performance, we shall see in the Section II.B that the \ac{CCSDS} \ac{LDPC} code design is particularly suited for application to satellite telecommand links.

The performance of a  turbo code introduced in \cite{Garello13} based on $16$-states component recursive convolutional codes is also provided (\ref{lab:turbo}). The turbo code shows superior performance with respect to binary \ac{LDPC} codes, down to low error rates. The code attains a \cer{-4} at almost $0.4$ dB from the \ac{RCB}. The code performance diverges remarkably from the \ac{RCB} at lower error rates, due to the relatively low code minimum distance.\footnote{In \cite{Jerkovits16:Short}, design improvements for the specific case of $(128,64)$ turbo codes have been presented, which are able to overcome the error floor issue down to \cer{-7}. The proposed design leverages on the use of tail-biting component codes together with a thorough interleaver search procedure.}

Results for both non-binary turbo (\ref{lab:tc256}) and \ac{LDPC} (\ref{lab:ldpc256}) codes are included in Figure \ref{fig:64128all}. Both codes have been constructed over a finite field of order $256$. The turbo code is based on memory-$1$ time-variant recursive convolutional codes \cite{Liva13:ShortTC}. The choice of memory-$1$ component codes enables the use of the \ac{FFT} to reduce complexity of their forward-backward decoding algorithm \cite{Berkmann02:dualtrellis}.  The non-binary \ac{LDPC} is based on an ultra-sparse parity-check matrix \cite{Poulliat08:BinImag}. Details on the code structure are provided in \cite{Liva12:CCSDS,CCSDS14,CCSDS15}. Both codes attain visible gains with respect to their binary counterparts, performing on the \ac{RCB} (and with $0.7$ dB from the normal approximation reference) down to low error rates (no floors down to \cer{-9} were observed in \cite{Liva12:CCSDS}).

For the block length considered in this comparison, a viable alternative to the use of codes with iterative decoders is provided by \ac{OSD}. Contrary to iterative decoding, \ac{OSD} \cite{Fossorier95:OSD} does not require any particular code structure, and hence can be applied to any (linear) block code. In Figure \ref{fig:64128all}, the performance of a $(128,64)$ extended \ac{BCH} code with minimum distance $22$ is displayed. The variant of \ac{OSD} used for the simulation is the one based on the identification of the most reliable basis. Test error patterns up to a maximum weight of $4$ have been used, resulting in a decoder list of $\approx 6.8 \times 10^{5}$ codewords. The \ac{BCH} code performance is close to the normal approximation benchmark, gaining $\approx 0.6$ dB over non-binary turbo and \ac{LDPC} codes at \cer{-4}. The same decoding algorithm has been applied to the binary image of the non-binary \ac{LDPC} code \ref{lab:ldpcosd}. Interestingly, a non-binary \ac{LDPC} \ac{CER} is almost indistinguishable from the one of the \ac{BCH} code, highlighting the sub-optimality of iterative decoding.
The error probability of a polar code under \ac{SC} decoding (\ref{lab:polar}) is included. A more appropriate comparison, able to fully exploit the potential of polar codes under list decoding, is a concatenation of an inner polar code with an outer high-rate error detection code as proposed in \cite{Tal15:ListPolar}. The error probability of the concatenation using a CRC-$7$ as an outer code   is shown (\ref{lab:polar_list_crc}). The polar code has parameters $(128,71)$, while the outer CRC code has generator polynomial $g(x)=x^7 + x^3 + 1$, leading to a code with dimension $64$. A list size of $32$ has been used in the simulation. The code outperforms all the competitors relying on iterative decoding algorithms, down to a \cer{-6}, where the code performance curve intersects the one of the non-binary turbo and \ac{LDPC} codes. It is anyhow expected that, by changing the design target for the polar code (resulting in a different set of frozen bits), a different trade-off between low and high \ac{SNR} performance can be achieved.

Finally, the \ac{CER} of three tail-biting convolutional codes has been included \cite{Stahl99:OptimumTB}. The first code (\ref{lab:tb8}) is based on a memory-$8$ encoder with generator polynomials (in octal form) given by $[515\,\, 677]$. The second code (\ref{lab:tb11}) is based on a memory-$11$ encoder with generator polynomials $[5537\,\,6131]$. The \ac{WAVA} algorithm has been used for decoding \cite{Fossorier03:WAVA}. The memory-$11$ convolutional code reaches the performance of the \ac{BCH} and \ac{LDPC} codes under \ac{OSD}. The memory-$8$ code loses $1$ dB at \cer{-6}, but still outperforms binary \ac{LDPC} and turbo codes over the whole simulation range. The third code (\ref{lab:tb14}) is based on a memory-$14$ encoder \cite{johannesson2015fundamentals} with generator polynomials (in octal form) given by $[75063 \,\, 56711]$. The code outperforms all other codes in Figure \ref{fig:64128all} (at the expense of a high decoding complexity due to the large number of states in the code trellis).

\subsection{The Elephant in the Room: Complexity}

In the comparison presented at the beginning of this Section, an important aspect has been (purposely) overlooked: the cost of decoding. The codes that perform close to the \ac{SPB} rely on relatively complex decoding algorithms. An exhaustive decoding complexity comparison would require a lengthy and rigorous analysis. Moreover, aspects that are not directly measurable in terms of algorithmic complexity (such as, for example, the probability vs. log-likelihood ratio domain form of the decoding algorithms) but still have large impact in hardware implementation can be difficultly compared. We provide next only a few qualitative remarks on complexity aspects for the decoding algorithms employed in the simulations.

\begin{theorem}[Binary vs. non-binary iterative decoding]
Binary iterative decoding for \ac{LDPC} and turbo codes can be efficiently performed in the logarithmic domain, with obvious benefits for finite precision (hardware) implementations. The belief propagation algorithm for the non-binary \ac{LDPC} and turbo codes presented in this manuscript is performed in the probability domain to allow for \ac{FFT}-based decoding at the check nodes \cite{Declerq03:FFT}. Thanks to the \ac{FFT}, complexity of iterative decoding is proportional to $q \log_2 q$ (being $q$ the field order), whereas the conventional iterative decoding complexity would scale with $q^2$. From an algorithmic complexity viewpoint, it has been estimated that the \ac{FFT}-based decoding of the $(128,64)$ non-binary \ac{LDPC} code is $\approx 64$ times larger than the one of (iterative decoding of) the \ac{CCSDS} \ac{LDPC} code \cite{CCSDS15}. Complexity reductions for non-binary \ac{LDPC} codes can be obtained by applying sub-optimum check node update rules, with various trade-offs between coding gain and decoding complexity \cite{Fossorier07:NB_DEC}.
\end{theorem}

\begin{theorem}[\ac{OSD} vs. non-binary iterative decoding]
For the code parameters adopted in this comparison, \ac{OSD} and non-binary belief propagation decoding over a finite field of order $256$ have similar decoding complexities, as documented in \cite{Baldi14:OSD}. However, the decoding complexity of \ac{OSD} scales less favorably with the block length than that of (non-binary) belief propagation (which is linear in the block length for a fixed iteration count and ensemble degree distributions pair). Hence, for larger block lengths \ac{OSD} may be considered impractical. Efficient \ac{OSD} variants have been introduced during the past decade, which may extend the range of interest for \ac{OSD} algorithms (see e.g. \cite{Fossorier04:BOX}).
\end{theorem}

\newcommand{\vecu}{\mathbf{u}}
\newcommand{\vecp}{\mathbf{p}}
\newcommand{\vecy}{\mathbf{y}}
\newcommand{\vecx}{\mathbf{x}}
\newcommand{\vecxML}{\mathbf{x}_{{\textsc{\tiny ML}}}}
\newcommand{\vecxOSD}{\mathbf{x}_{{\textsc{\tiny OSD}}}}
\newcommand{\vecuOSD}{\mathbf{u}_{{\textsc{\tiny OSD}}}}
\newcommand{\metML}{\Lambda_{{\textsc{\tiny ML}}}}
\newcommand{\metOSD}{\Lambda_{{\textsc{\tiny OSD}}}}

\newcommand{\pe}{P_{\mathsf{e}}}
\newcommand{\pue}{P_{\mathsf{ue}}}

\newcommand{\de}{\mathrm{d}}
\renewcommand{\dh}[1]{\mathrm{d_\textsc{\tiny h}}\left(#1\right)}

\newcommand{\code}{\mathcal{C}}
\newcommand{\listd}{\mathcal{L}}
\newcommand{\argmax}[1]{\underset{#1}{\mathrm{arg \, max}}\,}

\subsection{Error Detection}

Some of the algorithms used to decode the codes in Figure \ref{fig:64128all} are complete, i.e., the decoder output is always a codeword. Incomplete algorithms, such as belief propagation for \ac{LDPC} codes, may output an \emph{erasure}, i.e., the iterative decoder may converge to a decision that is not a (valid) codeword. Hence, while for complete decoders all error events are \emph{undetected}, incomplete ones provide the additional capability of discarding some decoder outputs when decoding does not succeed. In some applications, it is of paramount importance to deliver very low undetected error rates. This is the case, for instance, of telecommand systems, where wrong command sequences may be harmful. The \ac{CCSDS} \ac{LDPC} code has been designed with this objective in mind, trading part of the coding gain for a strong error detection capability \cite{Dolinar08:AngleMC}. Complete decoders, such as those based on \ac{OSD} and Viterbi decoding, may be used in such critical applications by adding an error detection mechanism. One possibility would be to include an outer error detection code. Nevertheless, in the short block length regime the introduced overhead might be unacceptable. In this context, a more appealing solution is provided by a post-decoding threshold test as proposed in \cite{Forney68:error_bounds}. Denote by $\mathbf{y}=(y_1,y_2,\ldots,y_n)$, with $\mathbf{x}+\mathbf{n}$, the \ac{bi-AWGN} channel output for a given transmitted codeword $\mathbf{x}$ ($\mathbf{n}$ is the noise contribution here). We refer to the conditional distribution of $\mathbf y$ given $\mathbf x$ as $p(\mathbf y| \mathbf x)$. We further denote the \ac{ML} decoder decision as
\[
\vecxML:=\argmax{\mathbf x \in \mathcal C} p(\mathbf y|\mathbf x).
\]
In \cite{Forney68:error_bounds} the metric
\begin{equation}
\metML(\vecy):=\frac{p(\vecy|\vecxML)}{\displaystyle\sum_{\substack{\vecx \in \code \\ \vecx \neq  \vecxML}} p(\vecy|\vecx)}\label{eq:Forney}
\end{equation}
was proposed
and it was proved that the rule for discarding the decoder decision given by the threshold test
\[
\metML(\vecy) < \exp(nT)
\]
is optimal in the sense on minimizing the undetected error probability for a given (overall) error probability. The metric \eqref{eq:Forney} is in general complex to compute (with some notable exceptions, see e.g. \cite{Hof09:CONV,Williamson14:ROVA}) due to evaluation of the denominator of \eqref{eq:Forney} (which requires a sum over all possible codewords) and to the need of the \ac{ML} decision $\vecxML$. In the case of \ac{OSD} (and of list decoders in general) an approximation of the metric \eqref{eq:Forney} can be easily obtained by summing the conditional distribution $p(\vecy|\vecx)$ over the codewords present in the list, only. The resulting metric would then be given by
\[
\metOSD(\vecy):=\frac{p(\vecy|\vecxOSD)}{\displaystyle\sum_{\substack{\vecx \in \listd \\ \vecx \neq  \vecxOSD}} p(\vecy|\vecx)}
\]
with
\begin{equation}
\vecxOSD:=\argmax{\vecx \in \listd} p(\vecy|\vecx). \label{eq:OSD}
\end{equation}
being $\listd$ the list produced by the \ac{OSD} algorithm. While the performance of the test based on the metric \eqref{eq:Forney} has been extensively studied (see e.g. \cite{Forney68:error_bounds,Hof10:Forney}) the authors are not aware of any attempt at analyzing the performance of the metric \eqref{eq:OSD}.

\section{Conclusions}

An overview of the recent efforts in the design and analysis of efficient error correcting codes for the short block length regime has been provided. A case study tailored to $(128,64)$ binary linear block codes has been used to discuss some of the trade-offs between coding gain and decoding complexity for some of the best know code/decoding schemes. The comparison, though incomplete, highlights some promising directions for the design of short and moderate-size block codes.

\section*{Acknowledgments}
The authors would like to thank Igal Sason for his valuable comments which helped to improve the review of finite-length performance benchmarks.



\begin{thebibliography}{10}
\providecommand{\url}[1]{#1}
\csname url@samestyle\endcsname
\providecommand{\newblock}{\relax}
\providecommand{\bibinfo}[2]{#2}
\providecommand{\BIBentrySTDinterwordspacing}{\spaceskip=0pt\relax}
\providecommand{\BIBentryALTinterwordstretchfactor}{4}
\providecommand{\BIBentryALTinterwordspacing}{\spaceskip=\fontdimen2\font plus
\BIBentryALTinterwordstretchfactor\fontdimen3\font minus
  \fontdimen4\font\relax}
\providecommand{\BIBforeignlanguage}[2]{{%
\expandafter\ifx\csname l@#1\endcsname\relax
\typeout{** WARNING: IEEEtran.bst: No hyphenation pattern has been}%
\typeout{** loaded for the language `#1'. Using the pattern for}%
\typeout{** the default language instead.}%
\else
\language=\csname l@#1\endcsname
\fi
#2}}
\providecommand{\BIBdecl}{\relax}
\BIBdecl

\bibitem{Shannon48}
C.~Shannon, ``A mathematical theory of communication,'' \emph{Bell System Tech.
  J.}, vol.~27, pp. 379--423, 623--656, Jul./Oct. 1948.

\bibitem{berlekamp74:key}
E.~R. Berlekamp, \emph{Key papers in the development of coding theory}.\hskip
  1em plus 0.5em minus 0.4em\relax IEEE Press, 1974.

\bibitem{Elias54:EFC}
P.~Elias, ``Error-free coding,'' \emph{Transactions of the IRE}, vol.~4, no.~4,
  pp. 29--37, September 1954.

\bibitem{Gallager63:LDPC}
R.~G. Gallager, \emph{Low-Density Parity-Check Codes}.\hskip 1em plus 0.5em
  minus 0.4em\relax Cambridge, MA, USA: M.I.T. Press, 1963.

\bibitem{Forney66:Concatenated}
G.~D. {Forney, Jr.}, \emph{Concatenated Codes}.\hskip 1em plus 0.5em minus
  0.4em\relax Cambridge, MA, USA: M.I.T. Press, 1966.

\bibitem{Costello07:PROC}
D.~J. {Costello, Jr.} and G.~D. {Forney, Jr.}, ``Channel coding: The road to
  channel capacity,'' \emph{Proceedings of the IEEE}, vol.~95, no.~6, pp.
  1150--1177, June 2007.

\bibitem{Berrou93:TC}
C.~Berrou, A.~Glavieux, and P.~Thitimajshima, ``Near {Shannon} limit
  error-correcting coding and decoding: Turbo-codes,'' in \emph{Proc. IEEE Int.
  Conf. Commun. (ICC)}, Geneva, Switzerland, May 1993.

\bibitem{MacKay99:LDPC}
D.~J.~C. MacKay, ``Good error-correcting codes based on very sparse matrices,''
  \emph{{IEEE} Trans. Inf. Theory}, vol.~45, no.~2, pp. 399--431, Mar 1999.

\bibitem{Richardson01:Design}
T.~Richardson, M.~Shokrollahi, and R.~Urbanke, ``Design of capacity-approaching
  irregular low-density parity-check codes,'' \emph{{IEEE} Trans. Inf. Theory},
  vol.~47, no.~2, pp. 619--637, Feb. 2001.

\bibitem{richardson01:capacity}
T.~Richardson and R.~Urbanke, ``The capacity of low-density parity-check codes
  under message-passing decoding,'' \emph{{IEEE} Trans. Inf. Theory}, vol.~47,
  no.~2, pp. 599--618, Feb. 2001.

\bibitem{Luby01:LDPC}
M.~Luby, M.~Mitzenmacher, M.~A. Shokrollahi, and D.~A. Spielman, ``{Improved
  Low-Density Parity-Check Codes Using Irregular Graphs},'' \emph{{IEEE} Trans.
  Inf. Theory}, vol.~47, no.~2, pp. 585--598, Feb. 2001.

\bibitem{Pfister05:capacity}
H.~D. Pfister, I.~Sason, and R.~Urbanke, ``Capacity-achieving ensembles for the
  binary erasure channel with bounded complexity,'' \emph{{IEEE} Trans. Inf.
  Theory}, vol.~51, no.~7, pp. 2352--2379, Jul. 2005.

\bibitem{Pfister07:ARA}
H.~D. Pfister and I.~Sason, ``Accumulate-repeat-accumulate codes:
  Capacity-achieving ensembles of systematic codes for the erasure channel with
  bounded complexity,'' \emph{{IEEE} Trans. Inf. Theory}, vol.~53, no.~6, pp.
  2088--2115, June 2007.

\bibitem{Arikan09:Polar}
E.~Arikan, ``Channel polarization: A method for constructing capacity-achieving
  codes for symmetric binary-input memoryless channels,'' \emph{{IEEE} Trans.
  Inf. Theory}, vol.~55, no.~7, pp. 3051--3073, July 2009.

\bibitem{Lentmaier10:CLDPC}
M.~Lentmaier, A.~Sridharan, D.~{Costello, Jr.}, and K.~Zigangirov, ``Iterative
  decoding threshold analysis for {LDPC} convolutional codes,'' \emph{{IEEE}
  Trans. Inf. Theory}, vol.~56, no.~10, pp. 5274--5289, Oct. 2010.

\bibitem{Kudekar11:CLDPC}
S.~Kudekar, T.~Richardson, and R.~Urbanke, ``Threshold saturation via spatial
  coupling: {W}hy convolutional {LDPC} ensembles perform so well over the
  {BEC},'' \emph{{IEEE} Trans. Inf. Theory}, vol.~57, no.~2, pp. 803 --834,
  Feb. 2011.

\bibitem{Decola11:CCSDS}
T.~de~Cola, E.~Paolini, G.~Liva, and G.~Calzolari, ``Reliability options for
  data communications in the future deep-space missions,'' \emph{Proc. IEEE},
  vol.~99, no.~11, pp. 2056--2074, Nov. 2011.

\bibitem{Boccardi14:MAG}
F.~Boccardi, R.~W. Heath, A.~Lozano, T.~L. Marzetta, and P.~Popovski, ``Five
  disruptive technology directions for {5G},'' \emph{{IEEE} Commun. Mag.},
  vol.~52, no.~2, pp. 74--80, Feb. 2014.

\bibitem{Paolini15:MAG}
E.~Paolini, C.~Stefanovic, G.~Liva, and P.~Popovski, ``Coded random access:
  applying codes on graphs to design random access protocols,'' \emph{{IEEE}
  Commun. Mag.}, vol.~53, no.~6, pp. 144--150, June 2015.

\bibitem{Durisi16:Short}
G.~Durisi, T.~Koch, and P.~Popovski, ``Towards massive, ultra-reliable, and
  low-latency wireless communications with short packets,'' \emph{Proc.
  {IEEE}}, 2016, to appear.

\bibitem{Richardson08:BOOK}
T.~Richardson and R.~Urbanke, \emph{Modern coding theory}.\hskip 1em plus 0.5em
  minus 0.4em\relax Cambridge University Press, 2008.

\bibitem{tenBrink01:EXIT}
S.~Ten~Brink, ``Convergence behavior of iteratively decoded parallel
  concatenated codes,'' \emph{{IEEE} Trans. Commun.}, vol.~49, no.~10, pp.
  1727--1737, Oct. 2001.

\bibitem{Sadjadpour00:shortTC}
H.~R. Sadjadpour, M.~Salehi, N.~J.~A. Sloane, and G.~Nebe, ``Interleaver design
  for short block length turbo codes,'' in \emph{IEEE International Conference
  on Communications}, vol.~2, 2000, pp. 628--632 vol.2.

\bibitem{Koetter03:Wheel}
C.~Radebaugh, C.~Powell, and R.~Koetter, ``{Wheel codes: Turbo-like codes on
  graphs of small order},'' in \emph{Proc. IEEE Inf. Theory Workshop (ITW)},
  Paris, France, Mar. 2003.

\bibitem{Ryan04:eIRA}
M.~Yang, W.~E. Ryan, and Y.~Li, ``Design of efficiently encodable
  moderate-length high-rate irregular {LDPC} codes,'' \emph{IEEE Transactions
  on Communications}, vol.~52, no.~4, pp. 564--571, April 2004.

\bibitem{Liva05:Tanner_Milcom}
G.~Liva and W.~Ryan, ``Short low-error-floor tanner codes with hamming nodes,''
  in \emph{Proc. IEEE Milcom}, Atlantic City, US, Oct. 2005.

\bibitem{Divsalar07:ShortProto}
D.~Divsalar, S.~Dolinar, and C.~Jones, ``Short protograph-based {LDPC} codes,''
  in \emph{Proc. IEEE Milcom}, Orlando, FL, USA, 2007, pp. 1--6.

\bibitem{Liva08:QC_GLDPC}
G.~Liva, W.~E. Ryan, and M.~Chiani, ``Quasi-cyclic generalized {LDPC} codes
  with low error floors,'' \emph{{IEEE} Trans. Commun.}, vol.~56, no.~1, pp.
  49--57, Jan. 2008.

\bibitem{Bocharova09:shortLDPC}
I.~E. Bocharova, B.~D. Kudryashov, R.~V. Satyukov, and S.~Stiglmayry, ``Short
  quasi-cyclic {LDPC} codes from convolutional codes,'' in \emph{IEEE
  International Symposium on Information Theory}, June 2009, pp. 551--555.

\bibitem{Divsalar14:ProtoRaptor}
T.-Y. Chen, K.~Vakilinia, D.~Divsalar, and R.~D. Wesel, ``{Protograph-Based
  Raptor-Like {LDPC} Codes},'' \emph{{IEEE} Trans. Commun.}, vol.~63, no.~5,
  pp. 1522--1532, May 2015.

\bibitem{Jerkovits16:Short}
T.~Jerkovits and B.~Matuz, ``Turbo code design for short blocks,'' in
  \emph{Proc. 7th Advanced Satellite Mobile Systems Conference}, Maiorca
  (Spain), September 2016.

\bibitem{Davey98:NonBinary}
M.~C. Davey and D.~MacKay, ``Low density parity-check codes over {GF(q)},''
  \emph{{IEEE} Commun. Lett.}, vol.~2, no.~6, pp. 165--167, Jun. 1998.

\bibitem{Berkmann00:Diss}
J.~Berkmann, \emph{{Iterative Decoding of Nonbinary Codes}}.\hskip 1em plus
  0.5em minus 0.4em\relax Munich, Germany: Ph.D. dissertation, Tech. Univ.
  M\"{u}nchen, 2000.

\bibitem{Berkmann02:dualtrellis}
J.~Berkmann and C.~Weiss, ``On dualizing trellis-based {APP} decoding
  algorithms,'' \emph{{IEEE} Trans. Commun.}, vol.~50, no.~11, pp. 1743--1757,
  Nov. 2002.

\bibitem{Poulliat08:BinImag}
C.~Poulliat, M.~Fossorier, and D.~Declercq, ``{Design of regular $(2,
  d_c)$-{LDPC} codes over {GF}(q) using their binary images},'' \emph{{IEEE}
  Trans. Commun.}, vol.~56, no.~10, pp. 1626--1635, 2008.

\bibitem{Venkiah08:RPEG}
A.~Venkiah, D.~Declercq, and C.~Poulliat, ``Design of cages with a randomized
  progressive edge-growth algorithm,'' \emph{{IEEE} Commun. Lett.}, vol.~12,
  no.~4, pp. 301--303, Apr. 2008.

\bibitem{Chen09:Hamilton}
W.~Chen, C.~Poulliat, and D.~Declercq, ``Structured high-girth non-binary cycle
  codes,'' in \emph{Asia-Pacific Conference on Communications (APCC)},
  Shanghai, China, Oct. 2009, pp. 462--466.

\bibitem{Costantini10:NBLDPC}
L.~Costantini, B.~Matuz, G.~Liva, E.~Paolini, and M.~Chiani, ``On the
  performance of moderate-length non-binary {LDPC} codes for space
  communications,'' in \emph{Proc. 5th Adv. Sat. Mobile Sys. Conf. (ASMS)},
  Cagliari, Italy, Sep. 2010.

\bibitem{kasai11:multiplicatively}
K.~Kasai, D.~Declercq, C.~Poulliat, and K.~Sakaniwa, ``Multiplicatively
  repeated nonbinary {LDPC} codes,'' \emph{{IEEE} Trans. Inf. Theory}, vol.~57,
  no.~10, pp. 6788--6795, Oct. 2011.

\bibitem{Liva12:IRAq}
G.~Liva, B.~Matuz, E.~Paolini, and M.~Chiani, ``Short non-binary {IRA} codes on
  large-girth {H}amiltonian graphs,'' in \emph{Proc. IEEE International Conf.
  on Commun. (ICC)}, Ottawa, Canada, Jun. 2012.

\bibitem{Divsalar12:NonBinaryShort}
B.~Y. Chang, D.~Divsalar, and L.~Dolecek, ``Non-binary protograph-based {LDPC}
  codes for short block-lengths,'' in \emph{Proc. IEEE Inf. Theory Workshop
  (ITW)}, Lausanne, Switzerland, Sep. 2012.

\bibitem{Liva13:ShortTC}
G.~Liva, E.~Paolini, B.~Matuz, S.~Scalise, and M.~Chiani, ``Short turbo codes
  over high order fields,'' \emph{{IEEE} Trans. Commun.}, vol.~61, no.~6, pp.
  2201--2211, June 2013.

\bibitem{Matuz13:LRNBTC}
B.~Matuz, G.~Liva, E.~Paolini, M.~Chiani, and G.~Bauch, ``Low-rate non-binary
  {LDPC} codes for coherent and blockwise non-coherent {AWGN} channels,''
  \emph{{IEEE} Trans. Commun.}, vol.~61, no.~10, pp. 4096--4107, October 2013.

\bibitem{Dolecek14:NBLDPCTIT}
L.~Dolecek, D.~Divsalar, Y.~Sun, and B.~Amiri, ``Non-binary protograph-based
  {LDPC} codes: Enumerators, analysis, and designs,'' \emph{{IEEE} Trans. Inf.
  Theory}, vol.~60, no.~7, pp. 3913--3941, July 2014.

\bibitem{Fossorier95:OSD}
M.~Fossorier and S.~Lin, ``Soft-decision decoding of linear block codes based
  on ordered statistics,'' \emph{{IEEE} Trans. Inf. Theory}, vol.~41, no.~5,
  pp. 1379--1396, Sep. 1995.

\bibitem{Fossorier04:BOX}
A.~Valembois and M.~Fossorier, ``Box and match techniques applied to
  soft-decision decoding,'' \emph{{IEEE} Trans. Inf. Theory}, vol.~50, no.~5,
  pp. 796--810, Dec. 2004.

\bibitem{Wu07:MRB}
Y.~Wu and C.~Hadjicostis, ``Soft-decision decoding using ordered recodings on
  the most reliable basis,'' \emph{{IEEE} Trans. Inf. Theory}, vol.~53, no.~2,
  pp. 829--836, Feb. 2007.

\bibitem{Liva14:OptProd}
G.~Liva, E.~Paolini, and M.~Chiani, ``On optimum decoding of certain product
  codes,'' \emph{{IEEE} Commun. Lett.}, vol.~18, no.~6, pp. 905--908, Jun.
  2014.

\bibitem{Tal15:ListPolar}
I.~Tal and A.~Vardy, ``List decoding of polar codes,'' \emph{{IEEE} Trans. Inf.
  Theory}, vol.~61, no.~5, pp. 2213--2226, May 2015.

\bibitem{Huber07:Multibase}
T.~Hehn, J.~B. Huber, S.~Laendner, and O.~Milenkovic, ``Multiple-bases
  belief-propagation for decoding of short block codes,'' in \emph{IEEE
  International Symposium on Information Theory}, Jun. 2007, pp. 311--315.

\bibitem{Huber10:Multibase}
T.~Hehn, J.~B. Huber, O.~Milenkovic, and S.~Laendner, ``Multiple-bases
  belief-propagation decoding of high-density cyclic codes,'' \emph{{IEEE}
  Trans. Commun.}, vol.~58, no.~1, pp. 1--8, Jan. 2010.

\bibitem{CCSDS}
\BIBentryALTinterwordspacing
Consultative Committee for Space Data Systems {(CCSDS)}. [Online]. Available:
  \url{http://www.ccsds.org}
\BIBentrySTDinterwordspacing

\bibitem{CCSDS14}
\emph{Next Generation Uplink}, Green Book, Issue 1, Consultative Committee for
  Space Data Systems {(CCSDS)} Report Concerning Space Data System Standards
  230.2-G-1, Jul. 2014.

\bibitem{shannon1967lower}
C.~E. Shannon, R.~G. Gallager, and E.~R. Berlekamp, ``Lower bounds to error
  probability for coding on discrete memoryless channels,'' \emph{Information
  and Control}, vol.~10, no.~1, pp. 65--103, 1967.

\bibitem{valembois2004sphere}
A.~Valembois and M.~P. Fossorier, ``Sphere-packing bounds revisited for
  moderate block lengths,'' \emph{{IEEE} Trans. Inf. Theory}, vol.~50, no.~12,
  pp. 2998--3014, Dec. 2004.

\bibitem{wiechman2008improved}
G.~Wiechman and I.~Sason, ``An improved sphere-packing bound for finite-length
  codes over symmetric memoryless channels,'' \emph{{IEEE} Trans. Inf. Theory},
  vol.~54, no.~5, pp. 1962--1990, May 2008.

\bibitem{Shannon59:SPB}
C.~Shannon, ``Probability of error for optimal codes in a {Gaussian} channel,''
  \emph{Bell System Tech. J.}, vol.~38, pp. 611--656, May 1959.

\bibitem{Gallager68:BOOK}
R.~Gallager, \emph{Information theory and reliable communication}.\hskip 1em
  plus 0.5em minus 0.4em\relax New York, NY, USA: Wiley, 1968.

\bibitem{Polyanskiy10:BOUNDS}
Y.~Polyanskiy, V.~Poor, and S.~Verdu, ``Channel coding rate in the finite
  blocklength regime,'' \emph{{IEEE} Trans. Inf. Theory}, vol.~56, no.~5, pp.
  2307--235, May 2010.

\bibitem{Dolinar98:BOUNDS}
S.~Dolinar, D.~Divsalar, and F.~Pollara, ``Code performance as a function of
  block size,'' Jet Propulsion Laboratory, Pasadena, CA, USA, TMO progress
  report 42-133, May 1998.

\bibitem{SasonShamai06:BOUNDS}
I.~Sason and S.~Shamai, ``Performance analysis of linear codes under
  maximum-likelihood decoding: A tutorial,'' \emph{Found. and Trends in Commun.
  and Inf. Theory}, vol.~3, no. 1--2, pp. 1--222, Jul. 2006.

\bibitem{SPECTRE}
\BIBentryALTinterwordspacing
Spectre: short packet communication toolbox. [Online]. Available:
  \url{https://sites.google.com/site/durisi/software}
\BIBentrySTDinterwordspacing

\bibitem{Thorpe03:PROTO}
J.~Thorpe, ``Low-density parity-check {(LDPC)} codes constructed from
  protographs,'' JPL IPN, Tech. Rep., Aug. 2003, 42-154.

\bibitem{Divsalar04:ARA}
A.~Abbasfar, K.~Yao, and D.~Disvalar, ``Accumulate repeat accumulate codes,''
  in \emph{Proc. IEEE Globecomm}, Dallas, Texas, Nov. 2004.

\bibitem{Divsalar09:ProtoJSAC}
D.~Divsalar, S.~Dolinar, C.~Jones, and K.~Andrews, ``Capacity-approaching
  protograph codes,'' \emph{IEEE JSAC}, vol.~27, no.~6, pp. 876--888, August
  2009.

\bibitem{Hu05:PEG}
X.~Hu, E.~Eleftheriou, and D.~Arnold, ``Regular and irregular progressive
  edge-growth tanner graphs,'' \emph{{IEEE} Trans. Inf. Theory}, vol.~51,
  no.~1, pp. 386--398, Jan. 2005.

\bibitem{Garello13}
M.~Baldi, M.~Bianchi, F.~Chiaraluce, R.~Garello, I.~Sanchez, and S.~Cioni,
  ``Advanced channel coding for space mission telecommand links,'' in
  \emph{IEEE VTC Fall}, Las Vegas, NV, USA, Sep. 2013, pp. 1--5.

\bibitem{Liva12:CCSDS}
G.~Liva, E.~Paolini, T.~D. Cola, and M.~Chiani, ``Codes on high-order fields
  for the {CCSDS} next generation uplink,'' in \emph{2012 6th Advanced
  Satellite Multimedia Systems Conference (ASMS) and 12th Signal Processing for
  Space Communications Workshop (SPSC)}, Sept 2012, pp. 44--48.

\bibitem{CCSDS15}
\emph{Short block length {LDPC} codes for {TC} synchronization and channel
  coding}, Orange Book, Consultative Committee for Space Data Systems {(CCSDS)}
  Experimental Specification 231.1-O-1, Apr. 2015.

\bibitem{Stahl99:OptimumTB}
P.~Stahl, J.~B. Anderson, and R.~Johannesson, ``Optimal and near-optimal
  encoders for short and moderate-length tail-biting trellises,'' \emph{{IEEE}
  Trans. Inf. Theory}, vol.~45, no.~7, pp. 2562--2571, Nov 1999.

\bibitem{Fossorier03:WAVA}
R.~Y. Shao, S.~Lin, and M.~P.~C. Fossorier, ``Two decoding algorithms for
  tailbiting codes,'' \emph{{IEEE} Trans. Commun.}, vol.~51, no.~10, pp.
  1658--1665, Oct 2003.

\bibitem{johannesson2015fundamentals}
R.~Johannesson and K.~S. Zigangirov, \emph{Fundamentals of convolutional
  coding}.\hskip 1em plus 0.5em minus 0.4em\relax John Wiley \& Sons, 2015.

\bibitem{Declerq03:FFT}
L.~Barnault and D.~Declercq, ``Fast decoding algorithm for {LDPC} over
  {GF}($2^q$),'' in \emph{Proc. IEEE Inf. Theory Workshop (ITW)}, Cergy,
  France, Mar. 2003, pp. 70--73.

\bibitem{Fossorier07:NB_DEC}
D.~Declercq and M.~Fossorier, ``Decoding algorithms for non-binary {LDPC} codes
  over {GF}$(q)$,'' \emph{{IEEE} Trans. Commun.}, vol.~55, no.~4, pp. 633--643,
  Apr. 2007.

\bibitem{Baldi14:OSD}
M.~Baldi, F.~Chiaraluce, N.~Maturo, G.~Liva, and E.~Paolini, ``A hybrid
  decoding scheme for short non-binary {LDPC} codes,'' \emph{{IEEE} Commun.
  Lett.}, vol.~18, no.~12, pp. 2093--2096, Dec 2014.

\bibitem{Dolinar08:AngleMC}
S.~Dolinar, K.~Andrews, F.~Pollara, and D.~Divsalar, ``Bounded angle iterative
  decoding of {LDPC} codes,'' in \emph{Proc. 2008 IEEE Milcom}, Nov. 2008, pp.
  1--6.

\bibitem{Forney68:error_bounds}
G.~D. {Forney, Jr.}, ``Exponential error bounds for erasure, list, and decision
  feedback schemes,'' \emph{{IEEE} Trans. Inf. Theory}, vol.~14, no.~2, pp.
  206--220, Mar. 1968.

\balance

\bibitem{Hof09:CONV}
E.~Hof, I.~Sason, and S.~{Shamai}, ``On optimal erasure and list decoding
  schemes of convolutional codes,'' in \emph{Proc. Tenth Int. Symp. Commun.
  Theory and Applications (ISCTA)}, Jul. 2009, pp. 6--10.

\bibitem{Williamson14:ROVA}
A.~R. Williamson, M.~J. Marshall, and R.~D. Wesel, ``Reliability-output
  decoding of tail-biting convolutional codes,'' \emph{{IEEE} Trans. Commun.},
  vol.~62, no.~6, pp. 1768--1778, June 2014.

\bibitem{Hof10:Forney}
E.~Hof, I.~Sason, and S.~Shamai, ``Performance bounds for erasure, list, and
  decision feedback schemes with linear block codes,'' \emph{{IEEE} Trans. Inf.
  Theory}, vol.~56, no.~8, pp. 3754--3778, Aug 2010.

\end{thebibliography}
\end{document}